\documentclass[10pt,conference]{IEEEtran}
\usepackage{times}
\usepackage{graphicx}
\usepackage{amsmath}
\usepackage{amssymb}
\usepackage{bm}

\begin{document}

\newtheorem{prop}{Proposition}
\newtheorem{lemm}[prop]{Lemma}

\title{
A Large Deviations Result for Aggregation of Independent Noisy Observations
}

\author{
\authorblockN{Tatsuto Murayama}
\authorblockA{NTT Communication Science Laboratories\\
NTT Corporation\\
Keihanna, Kyoto 619-0237, Japan\\
Email: murayama.tatsuto@labs.ntt.co.jp}
\and
\authorblockN{Peter Davis}
\authorblockA{NTT Communication Science Laboratories\\
NTT Corporation\\
Keihanna, Kyoto 619-0237, Japan\\
Email: peter.davis@labs.ntt.co.jp}
% \and
% \authorblockN{Muriel Medard}
% \authorblockA{Laboratory for Information and Decision Systems \\
% MIT \\
% Cambridge, MA \\
% Email: medard@mit.edu}
% \and
% \authorblockN{Amin Shokrollahi}
% \authorblockA{Lab. of Math. Algorithms \\
% EPFL\\
% Lausanne, Switzerland \\
% Email: amin.shokrollahi@epfl.ch}
% \and
% \authorblockN{Ram Zamir}
% \authorblockA{EE - Systems Dprt.\\
% Tel Aviv University\\
% Tel Aviv, Israel \\
% Email: zamir@eng.tau.ac.il }
}
% avoiding spaces at the end of the author lines is not a problem with
% conference papers because we don't use \thanks or \IEEEmembership
%Homer Simpson\authorrefmark{2},
%Montgomery Scott\authorrefmark{3} and
%Eldon Tyrell\authorrefmark{4}}
%Telephone: (800) 555--1212, Fax: (888) 555--1212}
%\authorblockA{\authorrefmark{4}Tyrell Inc., 123 Replicant Street, Los Angeles, California 90210--4321}}

\maketitle

\begin{abstract}
Sensing and aggregation of noisy observations should not be considered as separate issues.
The quality of collective estimation involves a difficult tradeoff between sensing quality which increases by increasing the number of sensors,
and aggregation quality which typically decreases if the number of sensors is too large.
We examine a strategy for optimal aggregation for an ensemble of independent sensors with constrained system capacity.
We show that in the large capacity limit
larger scale aggregation always outperforms smaller scale aggregation
at higher noise levels,
while below a critical value of noise,
there exist moderate scale aggregation levels
at which optimal estimation is realized.
\end{abstract}

\section{Introduction}
\label{sec:intr}

This letter presents results which give a new perspective
on the growing field of sensory data aggregation by clarifying fundamental principles of large-scale aggregation.
Examples of large scale aggregation of observations include astronomical observations~\cite{ryle1960slr},
biological sensing~\cite{franceschini1975sve},
early detection of natural disasters such as earthquakes,
tidal waves and floods~\cite{zschau2003ews}
and wireless sensor networks~\cite{kahn1999ncc}.
Errors in observations can be reduced by collecting observation data from more sensors.
However, collecting data from many sensors usually involves some cost
in terms of system resources, resulting in fundamental tradeoffs~\cite{akyildiz2002ssn}.
The theoretical understanding of these tradeoffs in natural and engineered systems is now a high priority.

An important fundamental problem in this field is the problem of aggregating independent observations of the same phenomenon
with a resource constraint. Previous works have analyzed the tradeoff behavior between aggregate data rate and sensing error from the fundamental view of information theory.  The analysis has been extended to include the situation where arbitrarily large numbers of samples can be collected by reducing the data aggregated from each sample using lossy data compression.
However, so far results have only been obtained for the fundamental information theoretic bounds with infinitely many sensors ~\cite{berger1996cpm,oohama1998rdf},
or specific situations in which the number of sensors is fixed ~\cite{gastpar2008uti}.
The previous works do not include the situation where the number of observations can be varied,
and thus the results are not sufficient to support our understanding
and design of real world systems.

In this paper we introduce a modification of the common basic model for data aggregation with compression which makes it more tractable and amenable to analysis when the number of sensors can vary.
Specifically, we consider independent decompression of each observation in
a discrete version of the CEO problem~\cite{berger1996cpm}.
We show that this model reveals a new property, the existence of noise threshold beyond which large scale aggregation is superior to lossless aggregation with no compression. This can be seen as a manifestation of
``more is different'' in sensor networks ~\cite{anderson1972md}.
Moreover, we show that universal results
for scaling behavior of collective estimation error
can be obtained by considering asymptotic behavior
when the system capacity diverges to infinity.

In this paper,
we consider a fundamental formulation of the problem with only one information
source and suppose that all sensors are symmetrical,
i.e.,
exchangeable with respect to their contributions to the final result of
aggregation.
This allows us to treat the problem in terms of the theory of large deviations.
The paper is divided into $5$ sections.
Section~\ref{sec:syst}
presents our system model.
Section~\ref{sec:stat}
briefly summarizes our main result.
The proof for the proposition,
however,
is postponed until the following Section~\ref{sec:larg}.
Discussions are given in Section~\ref{sec:disc}.

\section{System Model}
\label{sec:syst}

Now we start by introducing our system model for large-scale aggregation of independent noisy observations. Notice that we explicitly consider a capacity constraint. This section briefly summarizes the optimal strategy for the case of redundantly observing a Bernoulli$(1/2)$ sequence with very many sensors.

\subsection{Ensemble of Independent Sensors}

We consider that an observer
is interested in observing a purely random source $X$,
the state of which can be represented by a series of Ising variables $X_{\mu}$
and their realizations are explicitly denoted by the lower case letters
$x_{\mu}=\pm 1$.
We assume that this observer can not directly observe the source. 
Instead,
he deploys a collection of $L$ sensors,
labeled by an index $a$,
to independently observe the source and
report the results of their observations over a communication network.
Assuming a certain level of environmental noise,
the individual observations $Y_{\mu}(a)$
could be different for each sensor.
We define a common level of noise
$p \in [0,1/2)$
for our observations
\begin{align*}
\langle
\delta(X_\mu,-Y_\mu(a))
\rangle
=
p
%\label{eq:noise}
\end{align*}
with Kronecker's delta $\delta$,
where the braket
$\langle \ \cdot \ \rangle$
denotes the expectation of an argument.

Then we suppose that each sensor can compress (i.e., lossy encode)
if necessary,
its sensor readings
$\bm{y}(a)=(y_1(a),\cdots,y_M(a))$
into a codeword
$\bm{z}(a)=(z_1(a),\cdots,z_N(a))$
independently.
In this paper,
we assume that the codeword is represented by a series of Ising variables
$Z_\nu(a)$ and thus their realizations are restricted to $z_\nu(a)=\pm 1$
as well.
We further assume that the sensors themselves can not share any information
about their observations. 
That is,
they are not permitted to communicate with each other to decide what to send
beforehand.
As a result,
the observer must collect the $L$ codewords from all the sensors,
each of which separately encodes its own observations $y_\mu(a)$,
and use them to estimate the original $x_\mu$ for $\mu=1,\cdots,M$.
We assumed here that the lengths of the codewords are the same $N$,
so that all the sensors are identical with respect to
the ability of encoding their observations.
That is,
regardless of the sensor label,
the rate for the lossy encoding is given by $R=N/M$.
Therefore,
the load level of our network can be measured by the sum rate $LR$,
which should not be greater than the network capacity given by, say, $C$.
We assume that $C$ is a given integer, not a real,
in which case our argument will be greatly simplified.

If the sensors were able to share information about their observations before reporting to the observer, then they would be able to 
smooth out their independent environmental noises entirely
as the number of sensors $L$ diverges.
Then the observer can figure out all the realizations of $X_\mu$
if the network capacity $C$ exceeds $1$,
which is the entropy rate of the source $X$.
However,
if the mutual communications are prohibited,
there does not exist any finite value of $C$ 
for which even infinitely many sensors can transmit all the
information~\cite{berger1996cpm}.
Therefore,
our goal should be the semifaithful reconstruction of the original $x_\mu$
given the codewords $z_\nu(a)$
under a certain fidelity criterion.

\subsection{Exchangeable Sensor Ansatz}

Suppose that
$\hat{\bm{y}}(a)=(\hat{y}_1(a),\cdots,\hat{y}_M(a))$
be best reproductions for the observations obtained by using the codewords,
respectively.
Assume that the distortion between two sequences are always measured
by the Hamming distance per symbol.
Then it is easy to see that the distortion is given by,
in this case,
\begin{align*}
d(\bm{y}(a),\hat{\bm{y}}(a))
=(1/M)\sum_{\mu=1}^M\delta(y_\mu(a),\hat{y}_\mu(a))
\end{align*}
for $a=1,\cdots,L$.
Since we have exchangeable sensors as stated,
we can impose that
\begin{align*}
\langle d(Y(a),\hat{Y}(a)) \rangle=D
\end{align*}
for any given pairs.
With this Hamming distortion constraint,
the lower bound on the rate $R(D)$ required to describe a variable $Y_\mu(a)$
is given by
\begin{align*}
R(D)=1-H_2(D) \ ,
\end{align*}
where $H_2(D)$ denotes the binary entropy function~\cite{cover1991eit}.
This is called the rate distortion function
for the Bernoulli$(1/2)$ source.

The observer then collects all the transmitted information
$z_\nu(a)$
to calculate the estimate $\hat{x}_\mu$ for the $\mu$th
symbol of the unknown $\bm{x}=(x_1,\cdots,x_M)$.
To go further,
we now restrict ourselves to the case of
\begin{align*}
\langle \delta(Y_\mu(a),-\hat{Y}_\mu(a)) \rangle=D \ .
%\label{eq:distortion}
\end{align*}
That is,
every variable $\hat{Y}_\mu(a)$
in the reproductions is expected to have the same error probability $D$.
Notice also that the three variables
$X_\mu$, $Y_\mu(a)$, and $\hat{Y}_\mu(a)$
form a Markov chain,
when the best estimator for
$X_\mu$ is $\hat{Y}_\mu(a)$
if $0 \le p,D < 1/2$ holds.
Then it is straightforward to get,
independently,
\begin{align*}
\langle
\delta(X_\mu,-\hat{Y}_\mu(a))
\rangle
=
\rho \ ,
\end{align*}
where
\begin{align*}
\rho=p(1-D)+(1-p)D
\end{align*}
represents the combined error probability for replacing the original $x_\mu$
by the available symbols $\hat{y}_\mu(a)$.
In other words,
the error indicator function $\delta(X_\mu,-\hat{Y}_\mu(a))$
reduces to the Bernoulli random variable
that takes the value
$1$ with probability $\rho$ for $a=1,\cdots,L$.

\subsection{Bayes Optimal Estimator}

Now let us consider the most probable realization of $X_\mu$
given a set of evidences
$\hat{\bm{y}}_\mu=(\hat{y}_\mu(1),\cdots,\hat{y}_\mu(L))$.
Since $\delta(X_\mu,-\hat{Y}_\mu(a))$ obeys the Bernoulli statistics,
it is easy to see that
the majority vote procedure gives the best strategy~\cite{mackay2003iti}.
That is,
the optimal estimator should be a mapping
\begin{align*}
\hat{X}_\mu
=
\text{sgn}
\biggl\{
\sum_{a=1}^L
\hat{Y}_\mu(a)
\biggr\} \ .
\end{align*}
Then overall error probability for the estimate $\hat{x}_\mu$
is minimized.
The probability of getting more errors than $L/2$ out of $L$ Bernoulli
trials is given by
\begin{align*}
\begin{split}
&P(X_\mu \neq \hat{X}_\mu) \\
=&\langle \delta(X_\mu,-\hat{X}_\mu(a)) \rangle \\
=&\begin{cases}
\sum_{l=\frac{L+1}{2}}^{L}Q_\rho(l|L), & (\text{$L$ is odd}) \\
\sum_{l=\frac{L}{2}+1}^{L}Q_\rho(l|L)
+\frac{1}{2}Q_\rho(\frac{L}{2}|L) & (\text{$L$ is even})
\end{cases} \ ,
\end{split}
%\label{eq:general formula}
\end{align*}
where
\begin{align*}
Q_\rho(l | L)
=
{L \choose l} \rho^l (1-\rho)^{L-l}
\end{align*}
denotes the binomial distribution.
In principle,
we may choose whatever value of $L$ which is compatible with
the sum rate constraint of $LR \le C$.
To minimize the error probability for the estimator $\hat{X}_\mu$,
however,
we should use the largest possible value.
Hereafter
we assume that $L$ denotes the largest possible value.
In particular,
suppose that the sensors do not encode their observations.
Instead,
each sensor simply sends the whole information
of the noisy $Y_\mu(a)$.
Then,
the error probability for the estimator $\hat{X}_\mu$
reduces to
\begin{align*}
\begin{split}
&P(X_\mu \neq \hat{X}_\mu) \\
=&\begin{cases}
\sum_{l=\frac{C+1}{2}}^{C}Q_p(l|C), & (\text{$C$ is odd}) \\
\sum_{l=\frac{C}{2}+1}^{C}Q_p(l|C)
+\frac{1}{2}Q_p(\frac{C}{2}|C) & (\text{$C$ is even})
\end{cases} \ .
\end{split}
\end{align*}

\begin{figure}
\begin{center}
\includegraphics[scale=0.6,angle=0]{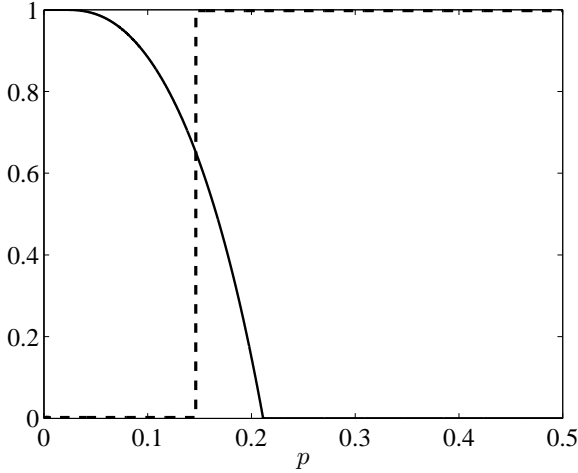}
\end{center}
\caption{
Optimal aggregation levels for ensemble of independent sensors in noisy environment. The $p$ is a given noise level. The solid line denotes the optimal data rate $R^*$ per sensor, which maximizes the exponential decay rate $I_p(R)$ of vanishing error probability with increase of system capacity $C$. For comparison, the dashed line represents the most pessimistic value $R^\dag$ which minimizes $I_p(R)$.
}
\label{fig:level}
\end{figure}

\section{Statement of Results}
\label{sec:stat}

An exact formula on the optimal data rate for individual sensors is presented in this section. By using the notion of large deviations an optimality measure for the data aggregation tasks is introduced. 
Numerical analysis of our exact result provides insights on the nature of large-scale aggregation in sensing systems, natural or engineered.

\subsection{Optimality Measure}

Assume that a network capacity $C$ is given.
Consider that the common data rate $R$
is first allocated to all the sensors.
The number of sensors $L$ is thus determined
as the maximum value of $L$ satisfying the sum rate constraint $RL \le C$.
In our system model,
it is obvious to say that
$P(X_{\mu} \neq \hat{X}_{\mu}) \to 0$
as $C \to \infty$.
As is shown in Section~\ref{sec:larg},
it is not hard to refine the above statement of convergence
and to prove that 
$P(X_{\mu} \neq \hat{X}_{\mu})$
decays to $0$ exponentially fast as $C \to \infty$.
By analogy with large deviation theory~\cite{ellis1985eld},
we define the exponential rate of decay by
\begin{align*}
I_p(R)
=
-\lim_{C \to \infty}
\frac{1}{C}\ln P(X_{\mu} \neq \hat{X}_{\mu})
\quad
(0<R \le 1) \ .
\end{align*}
The decay rate $I_p(R)$ describes the limiting behavior
of the system from a macroscopic level,
on which the rate $R$ could be used as a control parameter~\cite{murayama2009universal}.
The case of $R=1$ reduces to a naive aggregation scheme
in which the sensors just send their noisy observations
to the observer.
For this smallest aggregation,
we aggregate data from only $L=C$ sensors.
Hereafter,
we call this scheme the level-$1$ aggregation.
For a given $R>0$,
the level-$R$ aggregation is defined in which
every sensor encodes its observations at the rate of $R$ independently.
As an extension of the definition of $I_p(R)$ for $R>0$,
we could naturally define the level-$0$ decay rate as
\begin{align*}
I_p(0)
=
-\lim_{C \to \infty}
\frac{1}{C}
\lim_{R \to 0}
\ln P(X_{\mu} \neq \hat{X}_{\mu}) \ .
\end{align*}

\subsection{Large Deviations Result}

Assume that $D(R)$ denotes the distortion rate function,
which is the inverse function of $R(D)$.
Suppose that
\begin{align*}
\rho_p(R)=p(1-D(R))+(1-p)D(R) \ .
\end{align*}
Then, for $0 < R \le 1$,
the main result of this paper is given below.
\begin{prop}
We have
\begin{align}
I_p(R)
=
-\frac{1}{R}
\left\{
\ln 2+\frac{\ln \rho_p(R)}{2}+\frac{\ln (1-\rho_p(R))}{2}
\right\}
\ .
\label{eq:result}
\end{align}
\label{prop:result}
\end{prop}
The maximum of $I_p(R)$ is of great interest
from an engineering point of view.
That is,
we prefer larger values of $I_p(R)$.
Therefore,
we examine the optimal levels defined by
$R^*={\mathrm{argmax}}_{0 \le R \le 1}I_p(R)$.
The optimal aggregation,
for a given $p$,
is called the level-$R^*$ aggregation.

\subsection{Numerical Findings}

\begin{figure}
\begin{center}
\includegraphics[scale=0.6,angle=0]{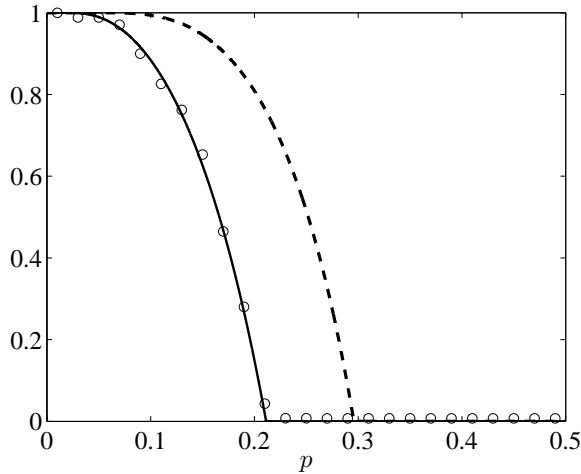}
\end{center}
\caption{
The solid line denotes the optimal data rate $R^*$ given by the large deviation analysis, while the dashed line represents the same value calculated by the Gaussian approximation in the prior work. The circles indicate the numerical experiments for $C=50$.
}
\label{fig:shift}
\end{figure}

We now examine the behavior of formula (\ref{eq:result})
which gives the optimal levels $R^*$ for the noise $p$.
As is seen in Fig.~\ref{fig:level},
the optimal aggregation scale diverges,
i.e.,
the optimal data rate $R^*$ per sensor diverges
for noise levels larger than the critical point $p_0=0.211$.
In this noisy region,
we want the system to be as large as possible.
The larger the system we have,
the smaller the error probability.
By definition,
the optimal aggregation is said to be level-$0$.
In contrast,
we can always find the non-zero optimal levels
below $p_0$.
In particular,
if the noise level is below $p_1=0.024$,
our investigations indicate that the level-$1$ aggregation is optimal.
Moderate aggregation levels could be optimal in the intermediate noise levels
between the two critical points.
It is also worth noticing that
the behavior of $R^*$ of $p$ is reminiscent of that of order parameters
at a continuous phase transition in statistical mechanics~\cite{monasson1999dcc}.
The analytical results presented here are also consistent with numerical
simulations for the system size $C=50$, as shown in Fig.~\ref{fig:shift}.

\begin{figure}
\begin{center}
\includegraphics[scale=0.6,angle=0]{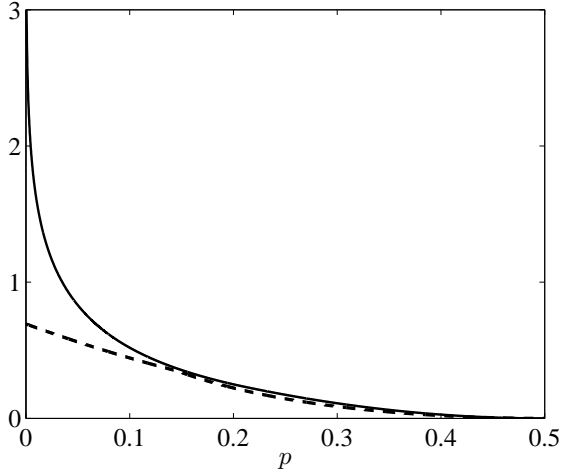}
\end{center}
\caption{
Maximum and minimum decay rates for vanishing error probability of final decision. The solid line denotes the largest decay rate $I_p(R^*)$ at the noise level $p$, which is given by the optimal data rate $R^*$ per sensor. For comparison, the dashed line represents the smallest decay rate $I_p(R^\dag)$ which is given by the most pessimistic value $R^\dag$.
}
\label{fig:rate}
\end{figure}

Since the optimal levels $R^*$ are unique values for each noise $p$,
we can plot the optimal decay rate $I_p(R^*)$ as is given in
Fig.~\ref{fig:rate}.
The optimal rate $I_p(R^*)$
describes the limiting behavior of the smallest error probability
$P(X_{\mu} \neq \hat{X}_{\mu})$
in terms of macroscopic variables.
Clearly,
it is a strongly decreasing function of the noise $p$.

\section{Analysis}
\label{sec:larg}

This section is devoted to present the large deviations analysis which gives Proposition~\ref{prop:result} and to describe briefly how it relates to the previous work by using the Gaussian approximation~\cite{murayama2009universal}. Numerical experiments support our recent result.

\subsection{Gaussian approximation}

For sufficiently large $L$,
the binomial distribution $Q_\rho(l|L)$
is well approximated by the Gaussian distribution
$\text{N}(L \rho,L \rho(1-\rho))$
with mean $L \rho$ and variance $L \rho(1-\rho)$~\cite{hays1981s}.
Changing the variable
\begin{align*}
s=\frac{l-L \rho}{\sqrt{L \rho(1-\rho)}}
\end{align*}
enables us to use a naive approximation to get
\begin{align}
P(X_\mu \neq \hat{X}_\mu)
\approx
\int_{\lambda_1}^{\lambda_2}
\frac{\text{d}s}{\sqrt{2 \pi}}
e^{-s^2/2} \ ,
\label{eq:definite integral}
\end{align}
where we denote,
respectively,
\begin{align*}
\lambda_1
=
\frac{1/2-\rho}{\sqrt{\rho(1-\rho)}}
\sqrt{L} \ ,
\quad
\lambda_2
=
\frac{1-\rho}{\sqrt{\rho(1-\rho)}}
\sqrt{L} \ .
\end{align*}
Since every sensor can achieve the optimal rate $R(D)$,
we may evaluate the number of sensors $L$ as $C/R(D)$.

Assume that $D(R)$ denotes the distortion rate function,
which is the inverse function of $R(D)$.
Suppose that $\alpha(p,R)=(1-2p)(1-2D(R))$ for $0\le p<1/2$
and $0<R \le 1$.
Together with an identity
\begin{align*}
\frac{1}{2}-\rho=(1-2p)
\left(
\frac{1}{2}-D
\right) \ ,
\end{align*}
we have estimated the rate function as
\begin{align*}
I_p(R)
=
\begin{cases}
\displaystyle
(1-2p)^2\ln 2
&
(R=0) \\
\displaystyle
\frac{\alpha(p,R)^2}{2R(1-\alpha(p,R))(1+\alpha(p,R))}
&
(0 < R \le 1)
\end{cases}
\ .
\end{align*}
However numerical evidence does not support the above formula, i.e., the Gaussian approximation (\ref{eq:definite integral}). This motivates us to apply the standard large deviation analysis, as shown below.

\subsection{Large Deviation Analysis}

Write the error indicator function $\delta(X_\mu,-\hat{Y}_\mu(a))$
as $Z_\mu(a)$.
For a given $\mu$,
this is a Bernoulli random variable
that takes the value
$1$ with probability
\begin{align*}
\rho=p(1-D)+(1-p)D
\end{align*}
for $a=1,\cdots,L$.
Consider the sample average defined to be
\begin{align*}
M_\mu=\frac{1}{L}\sum_{a=1}^LZ_\mu(a) \ .
\end{align*}
Since the expectation $\langle Z_\mu(a) \rangle=\rho$ is finite,
we know that $M_\mu$ is approaching $\rho$
by the law of large numbers.
However the value of interest is the error probability
$P(X_\mu \neq \hat{X}_\mu)$
for the majority vote procedure,
which is identical to
$P(M_\mu \ge 1/2)$.
For $0 \le \rho < 1/2$ and thus $|\rho-1/2|>0$,
the vanishing $P(M_\mu \ge 1/2)$ is called a large deviation probability.

Consider the rate function of $Z_\mu(a)$.
Since $Z_\mu(1)$, $Z_\mu(2)$, . . . , $Z_\mu(a)$
are the $L$ independent Bernoulli$(\rho)$ random variables,
the Legendre transform gives the rate function $I_\rho^{(1)}(z)$
for the sample average $M_\mu$ as
\begin{align*}
I_\rho^{(1)}(z)
=
z \ln \frac{z}{\rho}
+(1-z) \ln \frac{1-z}{1-\rho}
\end{align*}
for $0<z<1$ and $\infty$ otherwise~\cite{ellis1985eld}.
Since the number of sensors $L$ is given by $C/R$,
changing the variable
\begin{align*}
C=LR
\end{align*}
yields
\begin{align*}
I_{p,R}^{(1)}(z)
=
\frac{1}{R}
\left\{
z \ln \frac{z}{\rho_p(R)}
+(1-z) \ln \frac{1-z}{1-\rho_p(R)}
\right\} \ .
\end{align*}
Since the set $A_\mu=\{M_\mu \ge 1/2\}$
is closed and does not contain $\rho$,
the large deviation property tells that
\begin{align*}
\lim_{C \to \infty}
\frac{1}{C}
\ln P(A_\mu)
=
-\min_{z \in A_\mu}
I_{p,R}^{(1)}(z) \ .
\end{align*}
Then it is an easy matter to check that
\begin{align*}
\min_{z \in A_\mu}
I_{p,R}^{(1)}(z)
&=
I_{p,R}^{(1)}(1/2) \\
&=
-\frac{1}{R}
\left\{
\ln 2+\frac{\ln \rho_p(R)}{2}+\frac{\ln (1-\rho_p(R))}{2}
\right\}
\ .
\end{align*}
Write $I_p(R)=I_{p,R}^{(1)}(1/2)$ for the convenience.
For a given $R$ we conclude that 
\begin{align*}
I_p(R)
=
-\lim_{C \to \infty}
\frac{1}{C}
\ln P(A_\mu) \ .
\end{align*}
This completes the proof for Proposition~\ref{prop:result}.

\section{Discussion}
\label{sec:disc}

It has been shown that the optimal aggregation for an ensemble of independent sensors exhibits a critical behavior of the data rate per sensor $R=C/L$ with respect to the external noise level $p$. 
The simple analytic model shows that
in the high noise region beyond a critical value of noise $p_0$, the data rate $R$ should converge to zero in order to reduce collective estimation error.
This means that we should deploy very many sensors $L \gg C$ in the large $C$ limit.
In contrast, if the noise level is lower than the critical point, the data rate $R$ should take a positive value. In this case, the number of sensors scales as $L = \mathcal{O}(C)$.
Numerical evidence supports our large deviation analysis for the optimality measure.

\section*{Acknowledgment}

This work was in part supported by the
Ministry of Education, Culture, Sports, Science and Technology (MEXT) of Japan,
under the
Grant-in-Aid for Scientific Research on Priority Areas,
18079015.

\bibliographystyle{IEEEtran.bst}
\bibliography{IEEEabrv,isit2011}

\end{document}